# SERIOUS EDUCATIONAL REINFORCEMENT GAME IN PRESCHOOL

JUEGO SERIO DE REFUERZO EDUCATIVO EN ETAPA PRESCOLAR

Luis Ortegano, Esmitt Ramírez

*luis.ortegano.mora@gmail.com, esmitt.ramirez@ciens.ucv.ve*

## Resumen

Dentro del área de la Educación siempre se están buscando maneras de reforzar y ampliar las estrategias de enseñanzas para lograr que los estudiantes adquieran los conocimientos que se imparten en las asignaturas escolares, cursos de adiestramiento de personal y talleres. Los juegos serios forman parte de dichas estrategias de enseñanza, buscando reforzar o ampliar un conocimiento basado en un conjunto de reglas y normas de manera interactiva y entretenida. Actualmente, empleando la tecnología, es posible añadir imágenes y sonidos a los juegos serios bajo una plataforma digital. En este trabajo, se plantea una solución basada en juegos serios para reforzar el conocimiento de las matemáticas en la educación inicial dentro del plan de estudios de la República Bolivariana de Venezuela. Así, se ofrece una herramienta que se espera mejore las habilidades y conocimientos de los niños en un área específica, así como permitir hacer un seguimiento del progreso de los estudiantes recopilando la información de sus actividades. La recopilación se realiza empleando reportes que están disponibles a los docentes o facilitadores. Los reportes permiten ajustar la orientación y organización de los temas dictados en el aula, así como conocer el estado general del curso de prescolar. Las pruebas realizadas determinan el rendimiento computacional de la solución, así como el impacto en su aplicación en los niños y docentes. Se comprobó que la atención de los niños es amplia, logrando canalizar las actividades educativas de forma no directa en los niños.

## Abstract

In Education, is constant the searching of techniques to strengthen and extend the educational strategies in order to achieve that students get knowledge received in classroom, training personal courses and workshops. Serious games are part of that set of educational strategies to reinforce or extend the knowledge using a set of rules and policies in an interactive and amuse way. Nowadays, using technology, it is possible to add images and sounds into serious games under a digital platform. In this paper, we present a solution based on serious games to reinforce the mathematical knowledge in the preschool education following the study plan of the Bolivarian Republic of Venezuela. Thus, we offer a tool with expectation of improving skills and knowledge of children on a specific area, also allowing follow the student track getting information of the activities performed. This compilation is made performing reports which will be available for teachers and facilitators. Reports allow adjust orientation and organization of topics imparted in classroom, and to know the general overview of a preschool educational course. Test performed determine the computational performance of solution, also the impact on children and teachers. The wide attention of children was verified, accomplishing the educational tasks in an indirect way.

**Palabras Claves:** juegos serios; educación prescolar; matemáticas; enseñanza.

**Introducción**

La tecnología ha ido en constante crecimiento, ofreciendo grandes beneficios a gran parte de la población dentro del mundo moderno. Los dispositivos electrónicos han demostrado ser una herramienta de extrema utilidad para el ser humano moderno y se han adoptado como una herramienta primordial para el progreso. Desde un computador personal a dispositivos de menor capacidad de cómputo como los teléfonos y las tabletas, mantienen en contacto a las personas con sus roles de trabajo, vida social y hogar. Debido a la presencia de estos dispositivos de forma constantes en la mayoría de las aristas de la vida de un adulto en etapa laboral/productiva, su incorporación a niños y adolescentes es también creciente.

Según Williams (2014), para el año 2014 habían 1.8 billones de personas empleaban sus teléfonos inteligentes en su día a día. Igualmente, al menos el 70% de niños entre 11 y 12 años emplean su teléfono móvil, y al llegar a la edad de 14 años se incrementa a 90%. Según un estudio en Francia del año 2015 realizado por Solon (2016), el 58% de los niños por debajo de los 2 años ha empleado una tableta o teléfono inteligente. Esto demuestra un impacto inminente en un futuro cercano de estos dispositivos, haciendo abstracción de ciertos riesgos derivados (2014, Chapman y Pellicane). Entonces, al incorporar herramientas educativas en estos dispositivos, se llegará a los niños y adolescentes de forma directa o indirecta.

Actualmente es una práctica común que los padres acerquen dispositivos como teléfonos inteligentes y tabletas a los niños desde temprana edad en sus hogares y trabajos, para conseguir su diversión. De acuerdo con Kostyrka-Allchorne, Cooper y Simpson (2017), esto produce un estado de tranquilidad en el infante, que los hace permanecer tranquilos en un lugar, y facilitándole el acceso a múltiples canales de información y distracción. También, se busca contribuir al desarrollo o fortalecimiento de conocimientos del infante, aprovechando un tipo de aplicaciones que generen una sensación de disfrute, pero enfocados al desarrollo intelectual del niño y además de capacitarlo para el manejo de aplicaciones en dispositivos.

Una parte de estas aplicaciones son conocidas como juegos serios (2005, Michael y Chen), que son aplicaciones desarrolladas con la ayuda de un motor de videojuegos tal que generen más que diversión en el usuario final, estimulándolo a mejorar en algún sentido: concientizándolo en algún tema de interés, entrenándolo con situaciones similares a las cuales podría encontrarse envuelto, o fortaleciéndolo en los conocimientos adquiridos en aulas escolares.

Así, se encuentra una oportunidad de gran importancia en el emplear dispositivos móviles en las instituciones de educación inicial para que los niños den sus primeras incursiones a la tecnología empleando aplicaciones en dispositivos móviles. Aplicaciones que fortalezcan sus conocimientos a través de actividades orientadas a reforzar conocimientos específicos, además de realizar un control del progreso del niño en el transcurso de actividades realizadas con la aplicación.

En este trabajo se propone una solución que haga el uso de juegos serios para ayudar a impulsar a la tecnología dentro de la educación inicial en la República Bolivariana de Venezuela. Nuestra solución proporciona un conjunto de juegos serios diseñados para mejorar las habilidades y conocimientos de los niños en un área de conocimiento específica. Además, permite hacer seguimiento del progreso de los alumnos que hacen uso de la solución, recopilando sus resultados en todas las repeticiones de las actividades, y así generar reportes que estén disponibles para los docentes. Pudiendo emplear la aplicación por diversos alumnos y docentes, ajustados a la idea de perfiles de usuario.

Los reportes son una herramienta que pueden aportar drásticamente a la orientación y organización de los temas dictados. Estos permiten identificar las debilidades puntuales de los alumnos e identificar el estado general del curso en los distintos juegos que se proponen. En el transcurso del documento, se empleará el término niño para referirse a un infante, objetivo de nuestra investigación, sin diferenciación de género.

Este artículo describe brevemente el concepto de Juegos Serios en el ámbito de nuestra solución, para luego detallar los aspectos de nuestra propuesta, incluyendo cada módulo funcional. Posteriormente, se presentan detalles de la implementación del software realizado, para analizar un conjunto de experimentos. Finalmente, se presentan las conclusiones y trabajos a futuros planteados en esta investigación.

**1. Juegos Serios**

A primera vista, no resulta simple la definición de un juego serio basado en su expresión semántica. Michael y Chen (2005), definieron a los juegos serios como juegos que no tienen entretenimiento, disfrute y diversión como propósito principal. Su objetivo principal es enseñar, adiestrar, o entrenar a las personas, tales como personal militar, médico, estudiantes en todos los niveles e infantes. Los juegos serios pueden ser aplicados en el área de la salud, seguridad nacional, la ciencia, el gobierno, entrenamiento de personal corporativo, en estrategias de comunicación y educación en general.

En el campo militar, el uso de juegos serios es común para preparar al personal en situaciones como reconocimiento de entorno. Muchos de los juegos serios son configurables para poder brindarle al personal en entrenamiento un conocimiento tanto del área donde se va a desenvolverse, así como los posibles escenarios por las cuales podría transcurrir su operación.

Según el Serious Games Institute (2017), existen diversos videojuegos de venta masiva, que han sido modificados en sus configuraciones iniciales para ser empleados como juegos serios. Esto sucede ya que múltiples estudios muestran que las personas que hacen uso de videojuegos mejoran sus tiempos de respuestas a eventos rápidos y son capaces de seguir más elementos en movimiento que personas que no hacen uso de los videojuegos.

En el área de la Educación, se busca preparar a los estudiantes a adquirir un conjunto de habilidades, destrezas y conocimientos. En 1990, con el surgimiento de la capacidad multimedia de los computadores, aparecieron una serie de juegos educativos que se llamaron Edutainment (2007, Susi, Johanneson, y Backlund). El Edutainment se introdujo como actividades diseñadas para el entretenimiento y educación, concepto que está englobado en juegos serios. Sin embargo, tuvo poca aceptación debido a la baja calidad de los gráficos y al crecimiento global de Internet (2010, Breuer y Benten). Sin embargo, basados en estos, los juegos serios se han enfocado en propósitos específicos tornándose más aceptable para el refuerzo en áreas de matemática, lengua, espacial, lógica, reconocimiento, entre otras.

Un juego serio implementado en un computador se considera un videojuego. Así, los videojuegos deben ser clasificados de acuerdo al público al cual va dirigido, y acorde del contenido presentado. El ESRB (2017, Entertainment Software Rating Board) es un sistema de clasificación internacional de los videojuegos. Generalmente, los juegos serios tienen una clasificación Normal, las cuales comprenden EC (*Early Childhood*), E (*Everyone*), E 10+ (*Everyone 10 years and up*) y T (*Teen*).

Según los reportes presentados por Ambient Insight (2016, Adkins), los juegos serios tuvieron unas ganancias de $2.6 billones. Además, se plantea una categorización de los juegos serios y las simulaciones, proyectando para los próximos 5 años (año 2021) un crecimiento en los juegos serios en un 22.4%, y un incremento del 17% para los juegos que emplean aprendizaje basado en simulación.

Teniendo en cuenta el crecimiento de los juegos serios, países como China, EEUU, Corea del Sur y Japón, han invertido grandes cantidades de dinero en esta área con resultados exitosos (2016, Trevathan y otros). En Venezuela, existen algunos juegos serios enfocados en el área de educación tal como el videojuego El Sueño de Bolívar (2014, Prensa TeleSur), o un juego serio para la preservación de la fauna silvestre en peligro de extinción (2015, Lima, Torres y Ramírez). Del mismo modo, existen otras investigaciones en el área de Medicina (2013, Moreno y otros) (2016, Lupodia), o en Turismo (2013, Alfer). Así, en este artículo presentamos un juego serio para el refuerzo de algunos tópicos para educación de prescolar para dispositivos móviles, que es intuitivo, dinámico, que permite ser configurado y analizado por los tutores y representantes, y directamente asociado a la educación prescolar en Venezuela.

## 2. Nuestra Solución

En la Educación, existen diversas herramientas de apoyo dentro de las aulas de clases: libros, material didáctico, juegos de mesa, tabletas o computadores, entre otros. Sin embargo, generalmente fuera de las aulas de clases la cantidad de herramientas de apoyo son en menor cantidad. Sería ideal con una herramienta de apoyo para reforzar los conocimientos adquiridos, que pueda ser empleado tanto dentro como fuera del aula.

De esta forma, se plantea un juego serio para el refuerzo del contenido académico. Tomando como base la educación en el área de prescolar, basados en el plan de estudios de la República Bolivariana de Venezuela, específicamente en los contenidos del documento llamado *Sistema de Educación Inicial Bolivariana: Currículo y Orientaciones Metodológicas* (2007, Cenamec). En dicho documento se concibe la educación como un proceso social que comienza desde la gestación hasta cumplir los 6 años de edad, a través de la atención convencional y no convencional con la participación de la familia y comunidad.

Es claro que la computación educativa o computación en la educación se trata de intervenciones educativas, que requieren medir el impacto en los estudiantes. Todo esto empleando técnicas y herramientas de formación bajo la rigidez de un esquema educativo formal. Nuestra propuesta se remite a una herramienta de apoyo a la educación. Del mismo modo, se emplea una tableta como dispositivo de interacción con el niño y docente. Con este dispositivo se realizará todas las interacciones necesarias para que los niños aprendan a través de los juegos, y para que los docentes puedan obtener la evaluación de los niños a través de su desarrollo. El software solo se enfoca en utilizar los recursos locales del dispositivo para hacerlo independiente de la estructura de los planteles educativos (i.e. sin necesidad de conectividad a Internet).

La interacción está segmentada para dos usuarios finales, el principal que son los niños y lo más importante es que hagan uso de la herramienta para que fortalezcan sus conocimientos y además registren su progreso en la solución computacional. El otro tipo de usuario son los docentes que harán uso del módulo de juego para llevar un control de los avances de los niños a través de los reportes ofrecidos por la solución computacional.

## 2.1. Interacción del Niño

El registro de los niños debe ser lo más simple posible. Solo se requiere una fotografía del niño con el fin de simplificar su registro y su futuro reconocimiento de su usuario (ver Figura 1).

**Figura 1: Esquema general de los procesos de la solución**

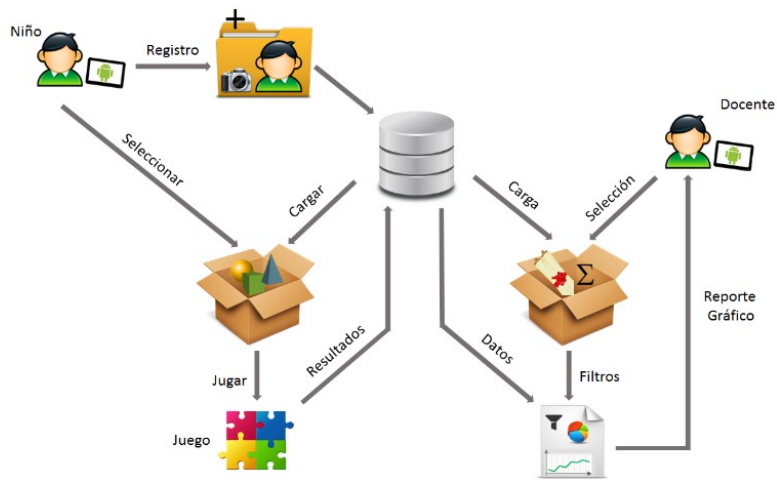

En la Figura 1, se muestra un flujo general de la solución. Se observa cómo actúa el niño en su proceso de registro, así como el flujo del proceso de seleccionar un juego. Nótese el flujo para el docente donde puede seleccionar un reporte, creando los filtros necesarios. Los datos generados por la solución se almacenan de forma permanente en una base de datos.

La solución se encargará de capturar una foto del niño y procesarla para crear su perfil y almacenar su información. Para los niños, seleccionar un juego consistirá en pocos pasos que los llevan rápidamente a hacer uso de los distintos juegos en la solución computacional. La Figura 2 muestra el ciclo que realiza un jugador desde seleccionar un juego y nivel hasta que finaliza el juego.

**Figura 2: Esquema general de los procesos de la solución**

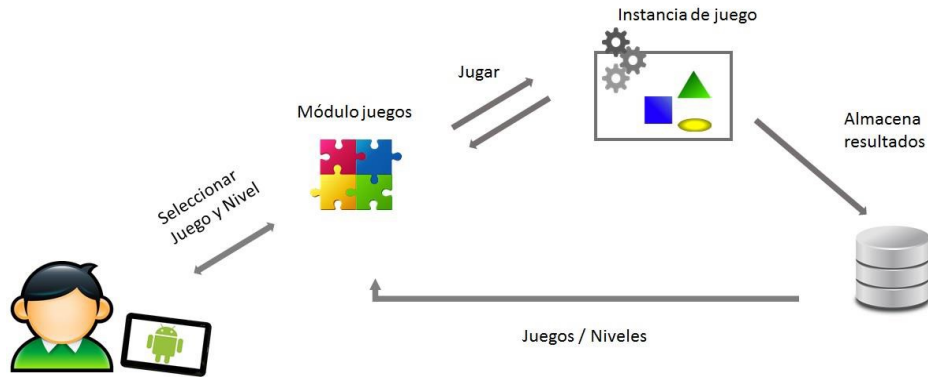

En principio, se presentan los niveles básicos al niño, a medida que supere otros juegos entonces podrá activar la aparición de nuevos juegos orientados a reforzar otras habilidades, o a una mayor dificultad. Los juegos tendrán una segunda categorización que está definida por niveles y representar la dificultad del juego seleccionado. Cada vez que se supera un nivel, se habilita un siguiente nivel con un grado de complejidad un poco mayor, aunque esto no es estrictamente necesario.

## 2.2. Interacción del Docente

Para el docente existe un módulo de reportes donde se presentan informes gráficos sobre el progreso del alumno en los distintos juegos. El objetivo de los reportes es crear recursos gráficos que fácilmente le indiquen al docente las fortalezas y debilidades de cada uno de los niños. De esta forma los docentes pueden ajustar sus actividades que se orienten a fortalecer las carencias detectadas en el curso.

En el módulo de reportes, existe una vía para complementar los datos de los niños con el propósito de hacer más fácil la identificación del niño para futuras ocasiones. Para poder visualizar los reportes el docente debe seleccionar inicialmente el reporte que desea generar de una lista desplegable, mostrando así una descripción de este e indicando información más detallada sobre este.

Cada reporte, tiene como finalidad mostrar un análisis sobre el comportamiento registrado del niño y así identificar sus habilidades. Existen ocasiones donde es necesario hacer una inspección, aplicando ciertos filtros sobre los reportes, para precisar otro tipo de información con más valor. Cada reporte posee sus propios filtros para inspeccionar en ciertos datos generados por un juego en particular, o condicionar la generación del reporte con los atributos comunes.

Luego de seleccionar los filtros el reporte se genera y muestra el resultado final tras haber consultado y procesado la información de los resultados de la base de datos.

## 2.3. Interacción de Juegos

Una de las principales características de nuestra propuesta, es la capacidad de incorporar nuevos juegos a la colección de manera sencilla, a través de un proceso de configuración de datos e implementación de rutinas de programación. Se dispone de un almacenamiento local para los datos correspondientes al juego, y los cuales pueden ser empleados por el módulo de juegos de manera que el niño pueda observar los mismos en la lista desplazable.

Adicionalmente, se deben configurar los niveles del juego para ser instanciados del mismo con los parámetros de configuración establecidos. En la Figura 3 se muestra un esquema del proceso necesario para la integración de un juego en la solución. Se muestra que es necesario insertar en base de datos los registros del juego y sus niveles correspondientes de manera que estén configurados correctamente. Esto genera una base de datos que la solución dará uso.

**Figura 3: Representación gráfica de los componentes para la integración de un nuevo juego**

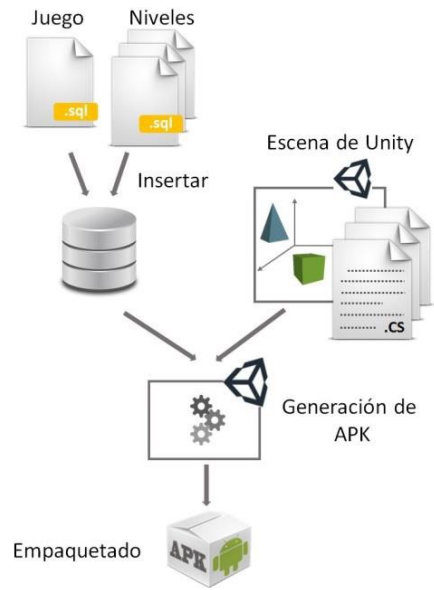

Un juego consta de una escena en (2018, Unity3d) donde se encuentran todos los objetos, dado que Unity3d permite añadir diversos elementos gráficos con características de audio, video e interactividad con el usuario. Este es un *engine* de juego (i.e. plataforma para el desarrollo de juegos) que permite desarrollar el juego como una aplicación.

Finalmente, una vez construido el juego, es necesario realizar un proceso de empaquetado o distribución para un dispositivo particular. En este proceso, Unity3d une los elementos que componen al juego, junto con la base de datos, y construye un solo archivo de extensión apk. Un archivo con extensión apk (*Android Application Package*) es una aplicación empaquetada para el sistema operativo Android (2018, Google).

Hasta este punto, el niño puede interactuar libremente en la aplicación del tipo APK que se encuentra instalado en el dispositivo. Posteriormente, el docente puede construir los reportes que miden el progreso de los niños que han usado la aplicación.

### 2.4. Integración de Reportes

El módulo de reportes comparte la filosofía de integración de los juegos, ya que, para medir nuevos juegos con distintos valores y resultados, existe la necesidad de tener reportes que se puedan adaptar a cada uno. Los datos y la definición de los reportes deben estar almacenados localmente en la base de datos para así poder consultarlos cuando el usuario ingrese al módulo de reportes.

Además de almacenar los reportes, se debe integrar un controlador que provea la creación y el manejo de los filtros que el usuario pueda utilizar para refinar los reportes. El procesamiento final del conjunto de los filtros genera un reporte para ser mostrado al usuario y obtener información acerca de los niños. Estos reportes quedan almacenados y pueden ser consultados en cualquier momento, seleccionados por la fecha cuando fue realizado los juegos por los niños.

## 3. Aspectos de Implementación

Desde la perspectiva de la implementación, se construyeron tres módulos principales: módulo de juegos, módulo de reportes y módulo de usuarios. El módulo de juegos y reportes, fueron explicados en la sección anterior, solo queda el módulo de usuarios. El módulo de usuarios permite al usuario registrarse a través de pasos sencillos, con el único requerimiento de capturar una foto para el registro. Luego de capturar la foto el usuario puede dirigirse directamente a hacer uso de los juegos que integra la solución.

El usuario está representado por una estructura de datos que reside en una base de datos local. Su tabla asociada tiene definido campos para el nombre y los apellidos del usuario, la modificación de estos valores se encuentra en el módulo de reportes para que el docente sea quien haga el registro de éstos. La imagen del usuario es almacenada en la memoria local del dispositivo identificándola.

Los juegos están definidos por una tabla en base de datos donde se almacena el nombre del juego y su descripción para que estén disponibles en el uso de los distintos módulos. Estos pueden estar ocultos para luego ser revelados al cumplir un nivel específico de otro juego al cual se relacione, de esta manera se provee un mecanismo para restringir otros juegos más complejos y darle la impresión al niño que está avanzando y obteniendo nuevos retos.

Los juegos se disponen visualmente en una lista con todos los juegos que el usuario actualmente puede seleccionar y jugar (ver tal como se muestra en la Figura 5. Los juegos que aún no haya descubierto permanecerán ocultos y desconocidos para el niño.

Figura 5: Vista previa del menú de selección de juegos

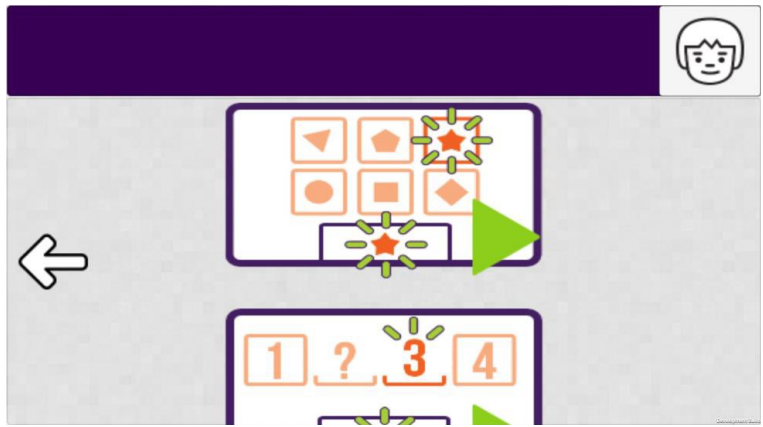

Los niveles están definidos por una tabla en base de datos que almacena el identificador único al juego que pertenece y el número del nivel que representa, además se dispone de un campo de texto donde se almacena la configuración del nivel en formato JSON (*JavaScript Object Notation*). El uso de JSON ofrece la libertad de modificar la estructura del objeto almacenado y se adapta a las posibles configuraciones que se necesite definir.

### 3.1. Herramientas

Se empleó el ambiente de desarrollo para Android Studio (2017, Google) el cual es el ambiente de desarrollo integrado oficial para la plataforma Android. Además, como se mencionó, se empleó el *engine* de Unity3d (2017, Unity3d) como herramienta.

Para el almacenamiento permanente en el dispositivo, se empleó SQLite (2017, Hipp), el cual es un sistema de gestión de base de datos relacional que es autocontenida. SQLite forma parte de la APK de forma directa, enlazándose de manera eficiente, lo cual hace que sea ideal en el consumo de recursos.

Para el despliegue de los reportes se empleó la biblioteca D3.js (2017, Bostock). Su nombre deriva de las siglas *Data-Driven Documents*, y es una biblioteca en Javascript que permite desplegar diagramas dinámicos e interactivos.

### 3.2. Selección de Juegos

Debido a que los juegos son componentes integrables a la solución computacional se han seleccionado dos habilidades mencionadas en el Currículo Nacional de la República Bolivariana de Venezuela para formar parte de los juegos pilotos de la aplicación. El currículo propone para el Sistema de Educación Inicial Bolivariana las áreas de aprendizaje y sus componentes, definiendo las habilidades y características que debe poseer los niños en esta etapa.

Para seleccionar juegos que respondan al perfil definido del subsistema de educación inicial se hizo referencia al libro de actividades de prescolar Matemática Inicial I (2010, Navarro) y Piruetas 1 (2014, Navarro) para identificar actividades que los niños realizan en este subsistema de educación. Además, se realizó un análisis de las actividades y resultaron dos juegos serios.

El primero de los juegos busca mejorar la habilidad de reconocimiento de formas geométricas a través de un proceso de identificación de parejas. Al niño se le presenta una forma geométrica y este debe realizar un proceso de búsqueda de la figura en un recuadro donde múltiples figuras son presentadas.

El segundo juego tiene como objetivo mejorar la identificación de los números y generar el sentido de orden de los números. La actividad que el niño debe realizar es poner en una serie de recuadros vacíos los números que la solución va disponiendo orden creciente.

### 4. Pruebas y Resultados

Con el objetivo de demostrar la eficacia de la solución, se requiere de un conjunto de pruebas en ambientes reales con estudiantes y docentes de la etapa prescolar. Primeramente, las pruebas se enfocaron en determinar el consumo de recursos durante la ejecución de la solución en un dispositivo. Luego, se efectuaron pruebas experimentales con niños en edades de etapa prescolar y, por último, se realizó un experimento con el grupo de docentes para completar la fase de pruebas.

## 4.1. Rendimiento

Unity3d integra una herramienta que ayuda a inspeccionar los juegos en desarrollo y detectar posibles puntos de mejoras al detectar consumos de hardware elevados. Aprovechando la disponibilidad de este instrumento, se utilizó como herramienta de medición de consumo de recursos en varios escenarios: creación y selección de un perfil; realización de un ciclo en un juego de la solución; y uso de los reportes disponibles en la solución.

Los valores medidos en el proceso de recolección de datos fue el uso de la CPU, la memoria consumida y la carga del componente de audio. Dichos valores fueron medidos en un LG Nexus 5 a una resolución de 1080x1920 píxeles, con Android Lollipop y un procesador Quad-core 2.3 GHz Krait 400.

Para la creación de un perfil, el consumo de recursos fue mínimo (no mayor a 40 Mb, y la aplicación total 87.9 Mb) y el juego mantiene un alto rendimiento, obteniendo resultado promedio en 100 fps (*frames per second*). En el proceso de pruebas de rendimiento para el uso de los juegos gráficamente, se pudo ver el incremento del consumo (16 a 17 ms por segundo por cada cuadro) por parte de las tareas mostrando que el proceso de total hacia mantener la tasa de cuadros por segundo estable de 60 fps.

Por su parte, la memoria consumida se mantuvo oscilante entre 80 Mb y 85 Mb, manteniendo un comportamiento similar en la fase anterior y sin representar un elevado costo para el sistema de prueba. Por último, la herramienta de monitoreo indicó que el uso de la CPU por parte de los procesos de audio en el sistema se mantiene en un consumo de un 0.5% del tiempo total. Esto representa una baja carga, registrando un consumo promedio de la aplicación en CPU un tiempo de 15.2 milisegundos.

Por último, cuando se accionan los reportes, el consumo de memoria fue más elevado que en otras instancias (125 Mb). Este comportamiento es predecible, debido a la manipulación de datos que se deben transferir desde la base de datos hacia el juego, para ser procesados. Sin embargo, el consumo se puede considerar despreciable en el dispositivo donde se efectuaron las pruebas.

**4.2. Aplicación en Niños**

Un conjunto de pruebas de la solución fue realizado con los niños en el Taller de Tareas dirigidas Tomasino, ubicado en Caracas, Altamira entre la Avenida 7 y 8. En esta institución se imparten tareas dirigidas de niños en etapa prescolar quienes hicieron uso de la solución.

Once (11) niños realizaron las pruebas y tienen edades comprendidas entre los cuatro (4) años y los siete (7) años, lo cual incluye a niños en primer nivel, segundo nivel y tercer nivel de la etapa prescolar. Para el resguardo y protección del menor, estas pruebas fueron realizadas previa autorización de la institución y con el consentimiento de las partes involucradas, entre ellas, la institución de tareas dirigidas y los padres o representantes de los niños que realizaron la prueba.

Al inicio de las pruebas se consultó a los participantes sobre su experiencia con dispositivos electrónicos como lo son computadores, teléfonos celulares, o tabletas preguntando: *¿El niño tiene experiencias previas con el uso de tabletas o teléfonos inteligentes?*

El resultado obtenido arrojó que un 90,90% de la población (10 niños) respondió positivamente a la pregunta. Esto indicó que los niños están familiarizados o han interactuado en gran parte con dispositivos electrónicos. También se preguntó a los niños que tipos de aplicaciones usaban y, en su mayoría indicaron que principalmente eran juegos de entretenimiento (i.e. clasificación ESRB tipo E). Ejemplo de estos juegos son Candy Crush (2017, King Ltd) y Cut The Rope (2017, ZeptoLab UK Limited), así como para ver videos bajo supervisión de sus representantes.

**Dinámica de Juego**

Los niños probaron dos juegos que se enfocan en el reconocimiento de figuras y ordenamiento de números. En ambos se desea emplear el menor tiempo posible. La primera parte de la evaluación consiste en identificar los aspectos de usabilidad de la aplicación y su aceptación por parte de los niños. Para ello, se deben realizar tres actividades:

1. Crear el usuario para comenzar el juego
2. Selección el usuario previamente creado
3. Seleccionar los juegos y completar varios niveles

La segunda parte de la evaluación consiste en la identificación del perfil, y la verificación del progreso en el uso de la solución. Para ello, se debe seleccionar un perfil creado previamente y, seleccionar los juegos y completar los niveles restantes.

**Resultados de la Primera Parte de la Evaluación**

El niño puede crear su perfil de forma intuitiva sin asistencia, o sin asistencia con una explicación sencilla, o con asistencia guiada en un registro de usuario. Basado en los resultados, el concepto de usuario y perfiles no se encuentra asimilado por parte de los niños. Un 81.81% de la población requirió un registro guiado, y el restante 18.19% lo hizo con una explicación sencilla. La asistencia consistió en dictar instrucciones al niño, para que realizará la actividad. Los niños de 6 y 7 años solo necesitaron una explicación sencilla (ver Figura 7).

**Figura 7: Asistencia a los niños, dándole una explicación sencilla**

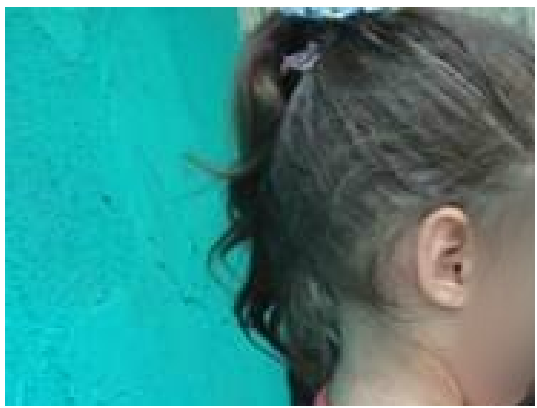

Posterior a la creación del perfil, se presenta un menú donde el niño debe seleccionar entre los juegos disponibles. El 72.73% pudo empezar un juego de forma intuitiva, el restante 27.27% necesito una explicación sencilla. Se nota que la selección de elementos en un menú es intuitiva. Los juegos presentan una explicación de cómo jugarlos de forma visual (i.e. sin texto en su explicación dentro del juego). El juego de reconocimiento de figuras, ver Figura 8, es similar a los juegos tradicionales que los niños están acostumbrados a ver en textos de ejercicios en sus centros de estudio. Casi todos los niños pudieron realizar este juego sin asistencia alguna, a excepción de 1 que requirió una explicación sencilla.

**Figura 8: Toma de pantalla del juego de reconocimiento de figuras**

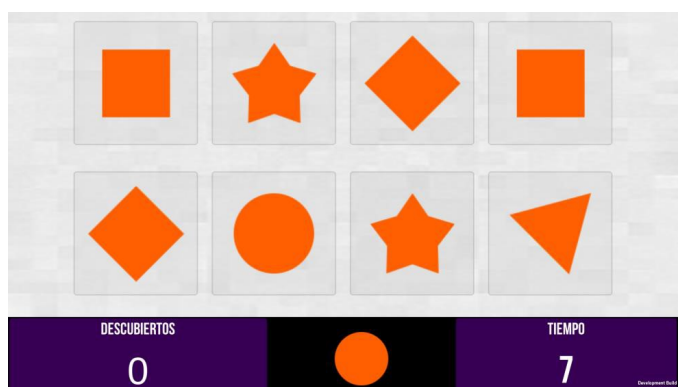

Por su parte, el juego de ordenamiento de números ocasionó distintas apreciaciones dado que no es un juego que pudiesen haber experimentado. El 72% de los niños requirió una explicación sencilla para comprender el objetivo del juego, el restante 18% lo pudo hacer sin explicación alguna.

**Resultados de la Segunda Parte de la Evaluación**

El objetivo de la segunda parte es comprobar que el juego sea de fácil comprensión en su dinámica (i.e. flujos de actividades). Esto inicia empleando la solución en la pantalla inicial donde se encuentra la selección de usuarios, y seguidamente debe realizar un juego y completar más niveles.

En cuanto a los perfiles, el 36% de los niños requirió una explicación del concepto de un perfil y asistencia, dado la dificultad de relacionar su perfil con sus actividades. El restante 64%, a pesar de no comprender el concepto, pudo identificar qué hacer luego de una explicación sencilla. En la Figura 9 se muestra un ejemplo de la selección de un perfil de usuario, mostrando una imagen del niño.

**Figura 9: Toma de pantalla para la selección del perfil del niño**

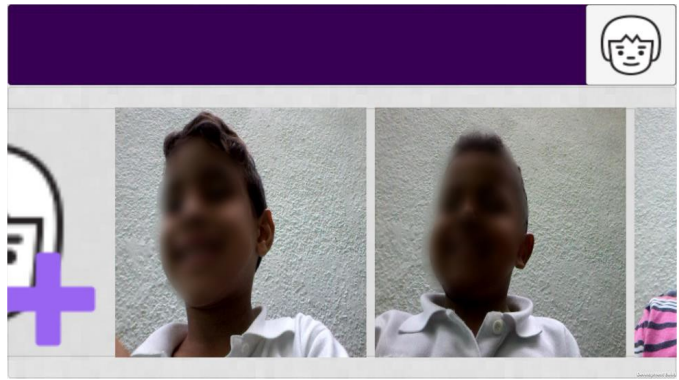

Luego de la selección del usuario del niño, se espera que el niño haga uso de los juegos, lo cual fue sin ninguna observación (todos lo realizaron).

**4.2. Aplicación en Docentes**

Esta fase de pruebas está orientada basado en el análisis de la experiencia que los docentes expresaron luego de hacer uso de la solución. En el área educativa de etapa prescolar los recursos electrónicos no son herramientas que abundan para el uso de los docentes, principalmente por las condiciones de los centros educativos (i.e. caso Venezuela - 2018). Las funcionalidades presentadas al docente fueron evaluadas iniciando con la sección de identificación de alumnos y luego haciendo uso de dos reportes integrados en la solución computacional.

**Identificación de Alumnos**

La funcionalidad de identificar a un alumno resultó sencilla por parte de la participante. Solo para completar los registros de varios usuarios, hubo la confusión al conocer a cuál alumno se estaba referenciando dado que no muestra la fotografía del alumno.

**Reportes**

En la selección de los reportes, se presentó una confusión cuando se mostró el reporte en el área de visualización. Esto fue por un mensaje que indicaba que aún el reporte no había sido seleccionado. Dicho mensaje se origina porque primero se debe hacer uso de los filtros para el uso del reporte. Esta instrucción fue indicada posteriormente, y se obtuvo el resultado deseado.

En cuanto a los resultados de los reportes, el juego del ordenamiento de números fue el que representó mayor reto para los alumnos (un 30% a 60% de victoria). Con este, se buscó obtener el reporte por rendimiento por día para identificar qué información aporta éste al docente en el análisis de las habilidades de los niños. Los comentarios del docente al observar la gráfica arrojaron que resultaba un poco compleja (ver Figura 10), entonces se tuvo que explicarles a los docentes su propósito. Este reporte aporta una medida para reconocer cuando un niño gana un juego y saber cuánto tiempo del invirtió, y cuántos desaciertos cometió. Y, cuando no completa un juego, se puede conocer que tan cerca estuvo de lograrlo, e indicar que tanto refuerzo necesita el niño.

Figura 10: Ejemplo de un reporte para un niño, obtenido por un docente

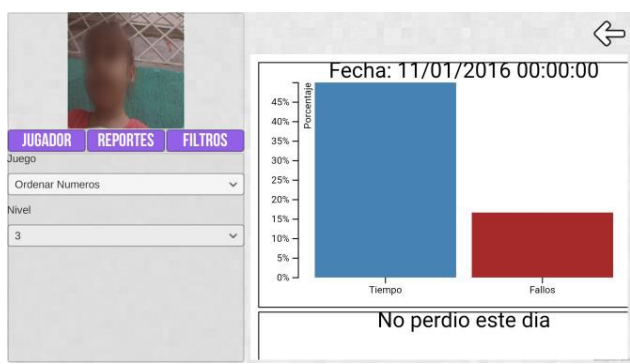

## 5. Conclusiones y Trabajo a Futuro

En este artículo se presenta una solución computacional destinada a los niños en etapa prescolar para el refuerzo educativo que puede ser empleada tanto dentro del aula, como fuera de ésta. Nuestra propuesta es un esquema multiusuario para permitir la utilización del dispositivo en un entorno escolar que resulta en un reto para los estudiantes como consecuencia de no conocer este tipo de sistemas. Sin embargo, esto no representa una limitante dado que demostraron tener la capacidad de asimilarlo con ayuda de un docente o representante, y con una explicación, poder manejar el concepto de usuario sin problema alguno.

Por otro lado, la integración de un motor de base de datos sin excesivo consumo de recursos computacionales aporta un fácil manejo de los datos internos del software. Además, SQLite es una herramienta que está siendo impulsada en el mercado por su alta confiabilidad y grandes prestaciones.

Los juegos seleccionados demostraron que nuestra propuesta puede atraer y mantener la atención de los niños aun cuando están diseñados específicamente para reforzar su conocimiento. Así, se espera que los juegos logren canalizar las actividades educativas para que, inconscientemente, los niños realicen ejercicios en beneficio de sus habilidades. La capacidad de incorporar nuevos juegos permite crear contenidos nuevos orientados y diseñados por entes educativos, con el fin de idear un conjunto de juegos que cubran contenidos educativos y usar la solución como un de libro de ejercicios, pero en este caso de manera digital.

El módulo de reportes creado para los docentes representa una importante característica, logrando ofrecer reportes gráficos para la fácil comprensión por parte de los padres y docentes. Este aporte logra dar a los supervisores un estatus de las habilidades puestas a prueba con la solución computacional y, con esta información poder planificar actividades que beneficien y se enfoquen en reforzar ciertos aspectos de los niños.

En un futuro, se propone construir un conjunto de iconografía e interacción que sean conocidos por los niños, es decir, que se encuentren dentro del rango de conocimiento a su corta edad. Al mismo tiempo, se plantea realizar un estudio intenso con la finalidad de demostrar en qué medida la solución se considera un refuerzo para la educación. Es importante aclarar que el proceso de certificación debe ser realizado por personas calificadas (e.g. profesionales de la docencia o psicólogos) para constatar la evolución de los niños.

**Agradecimientos**

**Referencias bibliográficas**